\documentclass[12pt]{iopart}

\usepackage{iopams}
\expandafter\let\csname equation*\endcsname\relax
\expandafter\let\csname endequation*\endcsname\relax
\usepackage{amsmath}
\usepackage{geometry}
\usepackage{graphicx}
\usepackage{subcaption}
\usepackage{multirow}
\usepackage{cite}
\usepackage{setspace}
\usepackage{color}
\usepackage{hyperref}
\hypersetup{urlcolor=blue, colorlinks=true}
\newcommand{\ACF}{\ensuremath{ \langle{\boldsymbol{ J}}(0)\otimes{\boldsymbol{J}(t)\rangle }}}

\begin{document}

\title[]{Comparison of the Green-Kubo and homogeneous non-equilibrium molecular dynamics methods for calculating thermal conductivity}

\author{B Dongre$^1$, T Wang$^2$, G K H Madsen$^1$}

 \address{$^1$ Institute of Materials Chemistry, TU Wien, A-1060 Vienna, Austria }
 \address{$^2$ Atomistic Modelling and Simulation, Interdisciplinary Centre for Advanced Materials Simulation (ICAMS), Ruhr-University Bochum, Germany}
 \eads{\mailto{georg.madsen@tuwien.ac.at}}
\begin{abstract}
Different molecular dynamics methods like the direct method, the Green-Kubo (GK) method and homogeneous non-equilibrium molecular dynamics (HNEMD) method have been widely used to calculate lattice thermal conductivity ($\kappa_\ell$). While the first two methods have been used and compared quite extensively, there is a lack of comparison of these methods with the HNEMD method. Focusing on the underlying computational parameters, we present a detailed comparison of the GK and HNEMD methods for both bulk and vacancy Si using the Stillinger-Weber potential. For the bulk calculations, we find both methods to perform well and yield $\kappa_\ell$ within acceptable uncertainties. In case of the vacancy calculations, HNEMD method has a slight advantage over the GK method as it becomes computationally cheaper for lower $\kappa_\ell$ values. This study could promote the application of HNEMD method in $\kappa_\ell$ calculations involving other lattice defects like nanovoids, dislocations, interfaces.  
\end{abstract}

%\submitto{\MSMSE}
\maketitle

\section{Introduction}
\label{intro}
Due to miniaturization of devices, increasing thermal loads and higher efficiency demands there is an ever-growing demand for materials with tailored thermal conductivities ($\kappa$) \cite{cahill2003nanoscale}. On one hand, devices like high power semiconductor electronic devices and lasers require high $\kappa$ for efficient heat removal which is critical to their performance. On the other hand, very low $\kappa$ is required in thermoelectric devices whose performance is inversely proportional to $\kappa$.

In semiconductors and insulators $\kappa$ is dominated by the lattice part $\kappa_\ell$. Classical molecular dynamics (MD) simulations are a powerful tool for calculating $\kappa_\ell$ due to their atomic resolution that helps to understand the underlying mechanisms leading to a specific $\kappa_\ell$. The conceptually simplest MD method for calculating $\kappa_\ell$ is the direct method\cite{schelling2002comparison, muller1997simple,mcgaughey2006phonon, stackhouse2010theoretical, turney2009predicting} where a temperature gradient is applied over a simulation box and the thermal conductivity is extracted from Fourier's law. However, the temperature gradients that must be applied are orders of magnitude larger than in experiments. Furthermore, the long mean free paths of the low frequency phonons make it difficult to converge the results \cite{sellan2010size}.

An alternate MD method for calculating $\kappa_\ell$ is the Green-Kubo (GK) method \cite{mcquarrie2000statistical}. The GK method applies linear response theory to the fluctuations of the heat current in a homogeneous equilibrium system. It requires integration of the auto-correlation function, which makes the method numerically challenging to apply. Another technique called homogeneous non-equilibrium molecular dynamics (HNEMD) circumvents the calculation of the auto-correlation function. The HNEMD method was initially proposed by Evans to calculate $\kappa_\ell$\cite{evans1982homogeneous}, who applied it to systems with pairwise interactions. The technique was subsequently applied to study systems with higher-order interactions among the particles\cite{berber2000unusually,lukes2007thermal,mandadapu2009homogeneous}.

To design materials with desired thermal conductivities, understanding the effects of defects is of prime importance. Defects like vacancies, dislocations, interfaces etc. are inevitably present in materials of technological interest. Such defects break the symmetry of the crystal structures and scatter the phonons thereby affecting the energy transported by phonons in solids. Vacancies in particular are very important because of their high concentrations at elevated temperatures. They act both as a large mass perturbation and perturb the bonding in the lattice and therefore strongly effect $\kappa_\ell$\cite{klemens1994thermal,wang2014atomistic,katcho2014effect, lee2011effects}. 

There has not been a systematic comparison of HNEMD with other methods in literature. In this work we fill this gap by
doing a systematic comparison of the GK and HNEMD methods in terms of their computational efficiencies and statistical errors when calculating $\kappa_\ell$. We perform calculations on both bulk-Si and Si with varying concentrations of vacancies using the Stillinger-Weber (SW) potential\cite{stillinger1985computer}.

\section{Background}
\label{sec:theory}
In order to calculate $\kappa_\ell$ with the GK and HNEMD methods, calculation of the heat current is essential. The heat current $\boldsymbol{J}$ is a vector quantity that characterizes the change with time of the spatial average of the local energy and is given as
\begin{equation}
\boldsymbol{J} = \frac{1}{\Omega} \frac{d}{dt} \sum_i \boldsymbol{r}_i(t) \epsilon_i(t)
\label{Eqn:J_siteenergy}
\end{equation}
where $\Omega$ is the volume of the system and $\epsilon_i$ and ${\bf r}_i$ are the total energy\cite{schelling2002comparison} and coordinate vector of atom $i$.
  
\subsection{The Green-Kubo method}
\label{subsec: GK}
The GK method\cite{green1954markoff,kubo1957statistical} is an equilibrium molecular dynamics approach that relates $\boldsymbol{J}$ to $\kappa_\ell$ via the fluctuation-dissipation theorem,
\begin{equation}
\boldsymbol{\kappa}_{\ell} = \frac{1}{k_BVT^2} \int \ACF dt 
\label{Eqn:kapa}
\end{equation}
where, $k_B$ is the Boltzmann constant, $V$ the volume of the system, $T$ the temperature and $\ACF$ the heat current auto-correlation function (HCACF). In general, $\kappa_\ell$ is a second-order tensor but in a material with cubic symmetry it reduces to a scalar.
The discretized form of the HCACF, Eq.~\eqref{Eqn:kapa}, is given as
\begin{equation}
\kappa_\ell = \frac{\Delta t}{k_BVT^2} \sum_{m=1}^M \sum_{n=1}^{N-m} \frac{J(m+n)J(n)}{N-m} 
\label{Eqn:kapa_discrt}
\end{equation}
where $\Delta t$ is the length of a single MD timestep and $M$ determines the length of the correlation time given by $M\Delta t$ over which the integration is done. $J(m+n)$ is the heat current at MD timestep $m+n$. It should be noted that the number of integration steps $M$ must be less than the total number of simulation steps $N$. 

Two potential errors can arise while calculating $\kappa_\ell$ via Eq.~\eqref{Eqn:kapa_discrt}. First of all, the correlation function $\ACF$ is usually calculated as a time-average from the data of a single simulation and using a finite averaging time $\tau_N = N\Delta t$ in Eq.~\eqref{Eqn:kapa_discrt} which leads to an averaging error.
Secondly, there is a truncation error related to $M$. Ideally $M$ should be so that the HCACF has decayed to zero\cite{schelling2002comparison}. Too small $M$ would underestimate $\kappa_\ell$ and a too large value would lead to large statistical errors as the heat current is dominated by noise after a certain correlation time \cite{howell2012comparison}. 

\subsection{Homogeneous non-equilibrium molecular dynamics.}
\label{subsec:HNEMD}
In HNEMD a fictitious force field is used to mimic the effect of a thermal gradient. It uses the linear response theory to calculate the transport coefficients in which the long-time ensemble average of the heat current vector, $\langle \boldsymbol{J}(t) \rangle$, for the resulting non-equilibrium system can be shown to be proportional to the external force field, $\boldsymbol{F}_{e}$, when the latter is sufficiently small\cite{evans1982homogeneous,evans1985nose}.

The detailed implementation of the HNEMD method for systems governed by three-body potentials can be found in Ref.~\cite{mandadapu2009homogeneous}. The equation to consider is,
\begin{equation}
\frac{\langle \boldsymbol{J}(t) \rangle}{VT}  = \left [\frac{1}{k_BVT^2} \int _0^\infty \ACF dt \right ] \boldsymbol{F}_e  = \boldsymbol{\kappa}_\ell \boldsymbol{F}_e
\label{Eqn:J_kFe}.
\end{equation} 
It shows a linear relationship between the external perturbation field $\boldsymbol{F}_\text{e}$ and the induced heat current $\boldsymbol{J}(t)$. The constant of proportionality ($\kappa_\ell$) is the GK formula for the heat transport coefficient tensor, Eq.~\eqref{Eqn:kapa}. In this way, the thermal conductivity can be obtained without explicitly calculating the auto-correlation functions. Hence, one can circumvent the  problems related to the calculation and integration of autocorrelation functions associated with the GK method described above. 

\section{Results and Discussion}
\label{sec:results}
We carried out all the simulations in silicon system on a 6$\times$6$\times$6 supercell containing 1728 atoms and with a lattice parameter of 5.431~$\mathrm{\AA}$  using the SW potential\cite{stillinger1985computer} and the LAMMPS package \cite{plimpton1995fast}. The choice of the parameters was based on previous finite size dependence studies done on silicon using MD techniques \cite{howell2012comparison, schelling2002comparison, che2000thermal}. In each simulation, after the velocity initialization the system was equilibrated under zero pressure at 1000~K (NPT ensemble).

\subsection{Bulk conductivity}
\label{subsec:bulk}
From {\bf the GK method}  the thermal conductivity was obtained through Eq.~\eqref{Eqn:kapa_discrt}. The HCACF was sampled at every timestep for better accuracy of the correlation integral. The simulations were then run in the NVE ensemble for the calculation and sampling of the HCACF. A time step of $\Delta t = 1.0$~fs was chosen to ensure long-time stability of the HCACF.

The averaging error can be minimized either by performing one simulation for a very large number of time steps or by carrying out multiple independent simulations for smaller time durations. We have applied the latter technique in which the simulations can be run in parallel and we collected a total of $60$~ns of data comprising of 5 independent simulations of $N=12$~ns ($12\times10^6$ timesteps) each. These independent simulations vary only with respect to the random seed provided for generating the initial atomic velocities\cite{howell2012comparison, schelling2002comparison}. The total correlation length for a single simulation was 300~ps ($3\times10^5$ timesteps).  

\begin{figure}[t!]
%\hspace*{-2.3cm}
	\begin{subfigure}[t]{0.5\textwidth}
		\centering
		\includegraphics[width=\textwidth]{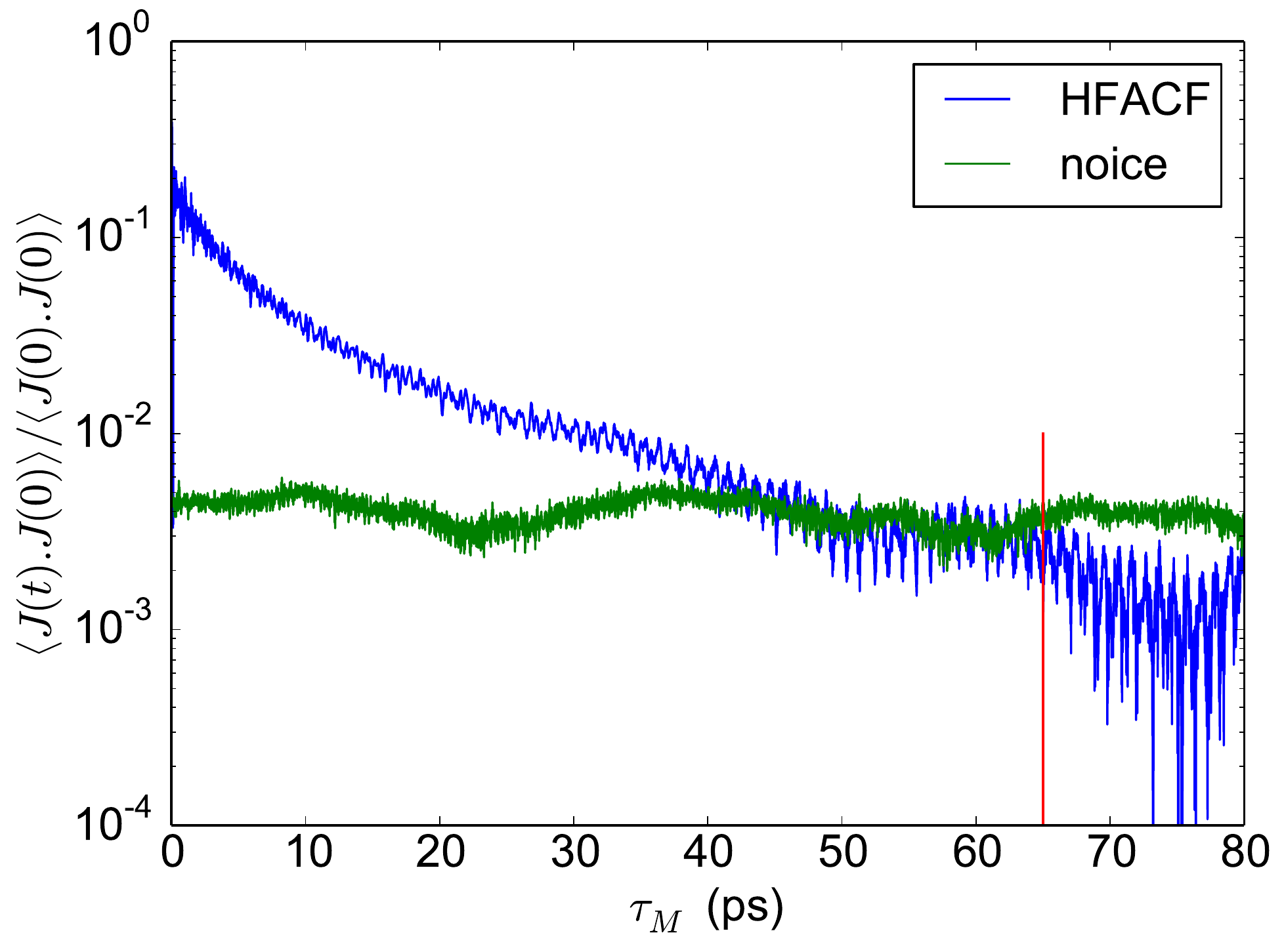}
		\caption{}
		\label{fig:ACF_Si}
	\end{subfigure}
	\begin{subfigure}[t]{0.5\textwidth}
		\centering
		\includegraphics[width=\textwidth]{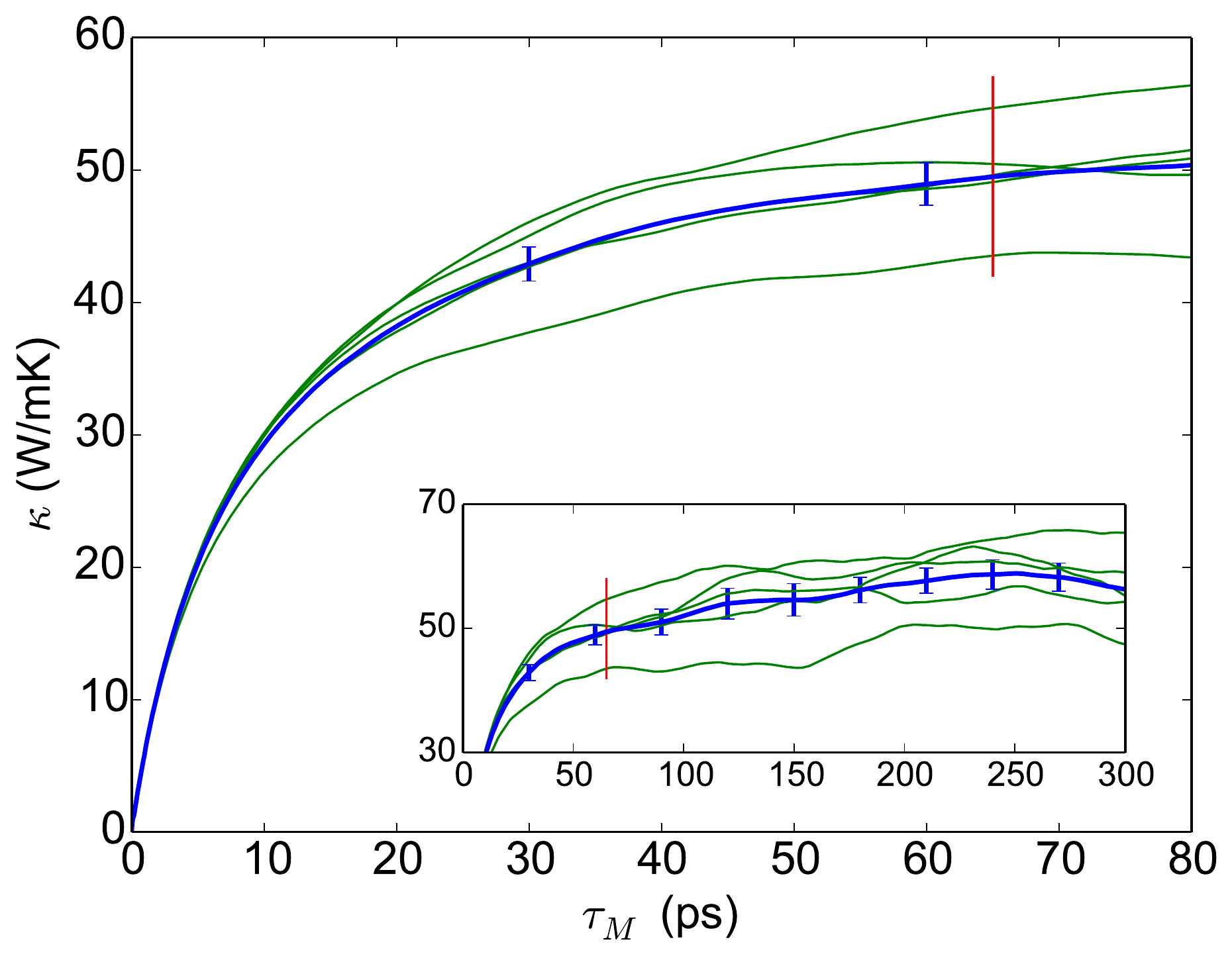}
		\caption{}
		\label{fig:Kappa_Si}
	\end{subfigure}
\caption{(a) Representative HCACF for Si (SW potential) as a function of $\tau_M=M\Delta t$ , normalized to the value at $t = 0$ (Blue line). (b) Thermal conductivity of Si calculated by the direct integration of the HCACF, with the redline showing $\tau_c$. The blue line depicts the averaged thermal conductivity over 5 independent simulations (green lines). Also shown in the inset is $\kappa_\ell$ calculated for the total correlation time of 300~ps and error bars can be seen growing rapidly after 65~ps (red line) because of the calculations being dominated by noise.}
\label{fig:Si_GK_HCACF_kapa}
\end{figure}

In order to minimize the truncation error, we employed the idea of a maximum significant correlation time $(\tau_c)$~\cite{jones2012adaptive} to determine an optimum $M$. Fig.~\ref{fig:ACF_Si} shows the ensemble averaged HCACF together with the noise (green line) which is estimated as the root-mean-square of the heat current cross-correlation functions \cite{howell2012comparison},
\begin{equation}
\zeta = \sqrt{\frac{\left \langle J_x(t)J_y(0) \right \rangle^2 + \left \langle J_y(t)J_z(0) \right \rangle^2 + \left \langle J_z(t)J_x(0) \right \rangle^2}{3}}.
\label{Eqn:W_noise}
\end{equation}
$\tau_c$ is the time after which the noise crosses and becomes larger than the HCACF and  is shown as a red line in Fig.~\ref{fig:ACF_Si}. The HCACF can be seen to decay and at $\tau_M \equiv \tau_c \simeq 65$~ps (red line) the noise  overtakes the HCACF.

Next, $\kappa_\ell$ was calculated by direct integration of the HCACF. Fig.~\ref{fig:Kappa_Si} shows the value of $\kappa_\ell$ vs the correlation time. The integration till $\tau_c$=65~ps gives a value of 49.6$\pm$1.5~W/(mK). This value is in good agreement to the other GK calculations $53.3\pm5.2$ W/(mK) \cite{jones2012adaptive, mcgaughey2006phonon}, lower than but consistent with $66\pm16$ W/(mK) \cite{schelling2002comparison}. It is also slightly lower than $53 \pm 3$ W/(mK) reported by Howell \textit{et.~al.~}\cite{howell2012comparison} because they calculated $\kappa_\ell$ by integrating till $\infty$ an exponential function fitted to the values of $\kappa_\ell$ averaged over five independent samples. They mentioned that the contribution to the total value of $\kappa_\ell$ by correlation times greater than $\tau_c$ is $<5\%$. Recently, Jones \textit{et.~al.~}\cite{jones2012adaptive} proved that the relative error in the estimate of $\kappa_\ell$ is bounded by the ratio of maximum significant correlation time $(\tau_c)$ and the total simulation time $K$ as $2\sqrt{\tau_c/K}$, which in our case reduces to $\approx \sqrt{M/N}$. This limits the relative error to be less than 6.5\%, which is consistent with our results.

\begin{figure}[tbp]
	\centering
		\includegraphics[scale=.6]{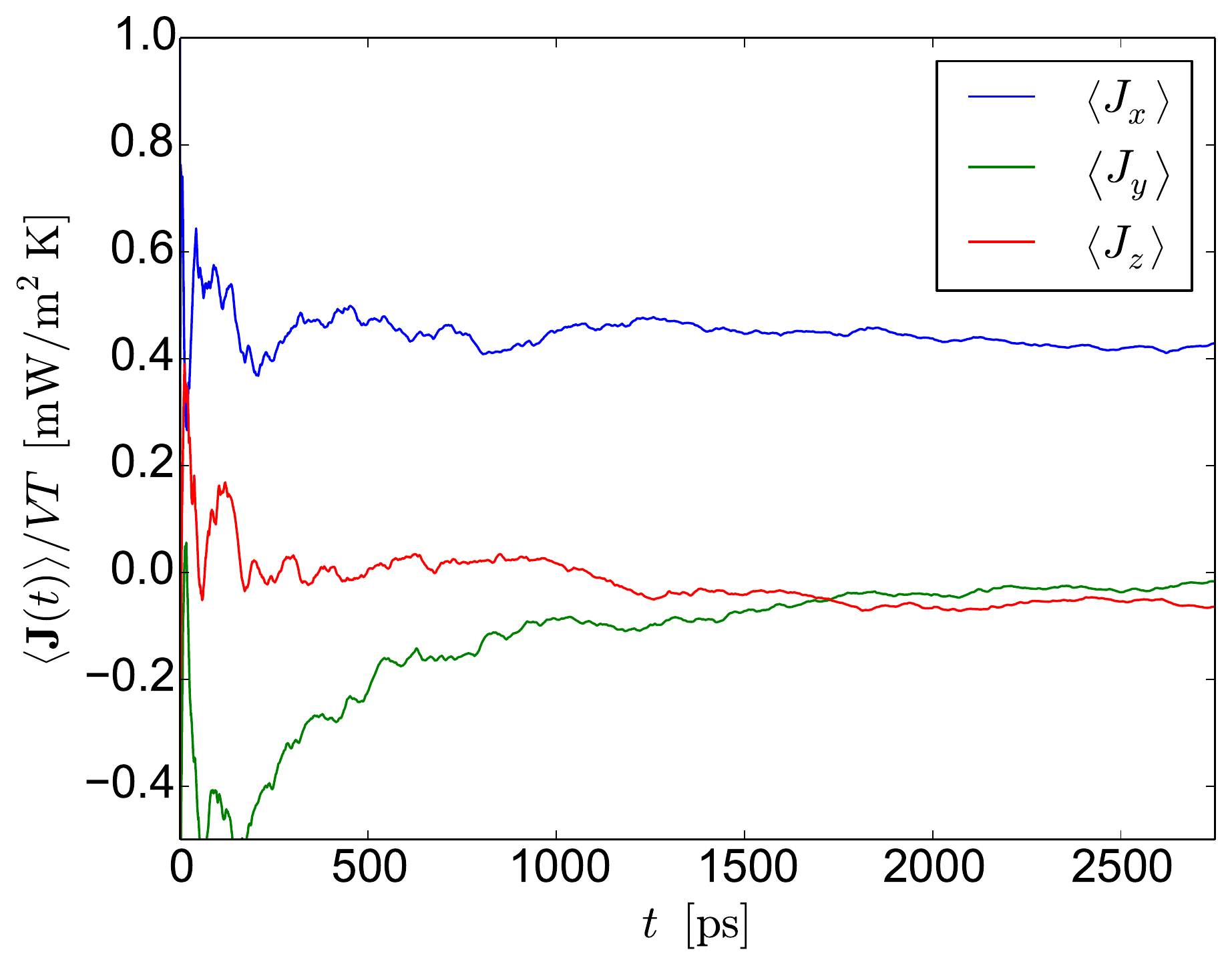}
		%\rule{35em}{0.5pt}
	\caption[J vs t for bulk Si]{The running average of components of the heat current $\boldsymbol{J}$ in the $x$, $y$ and $z$ directions for Si with $F_{e} = 8 \times 10^{-5} \mathrm{\AA^{-1}}$ applied in the $x$ direction.}
	\label{fig:Si_bulk_JvsT}
\end{figure}

\begin{figure}[t!]
%\hspace*{-2.3cm}
	\begin{subfigure}[t]{0.5\textwidth}
		\centering
		\includegraphics[width=\textwidth]{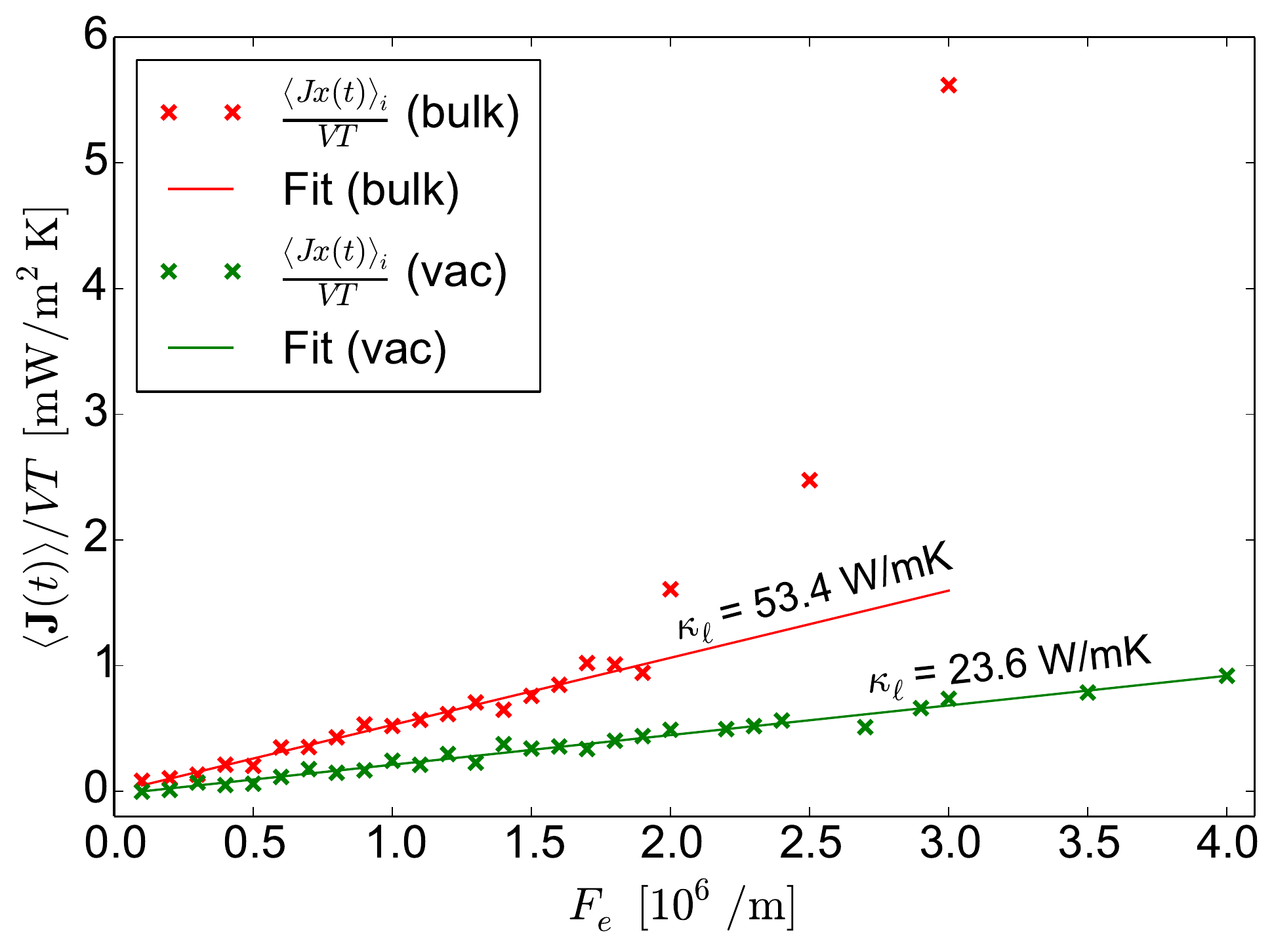}
		\caption{}
		\label{fig:Si_bulk_vac_JvsF}
	\end{subfigure}
	\begin{subfigure}[t]{0.5\textwidth}
		\centering
		\includegraphics[width=\textwidth]{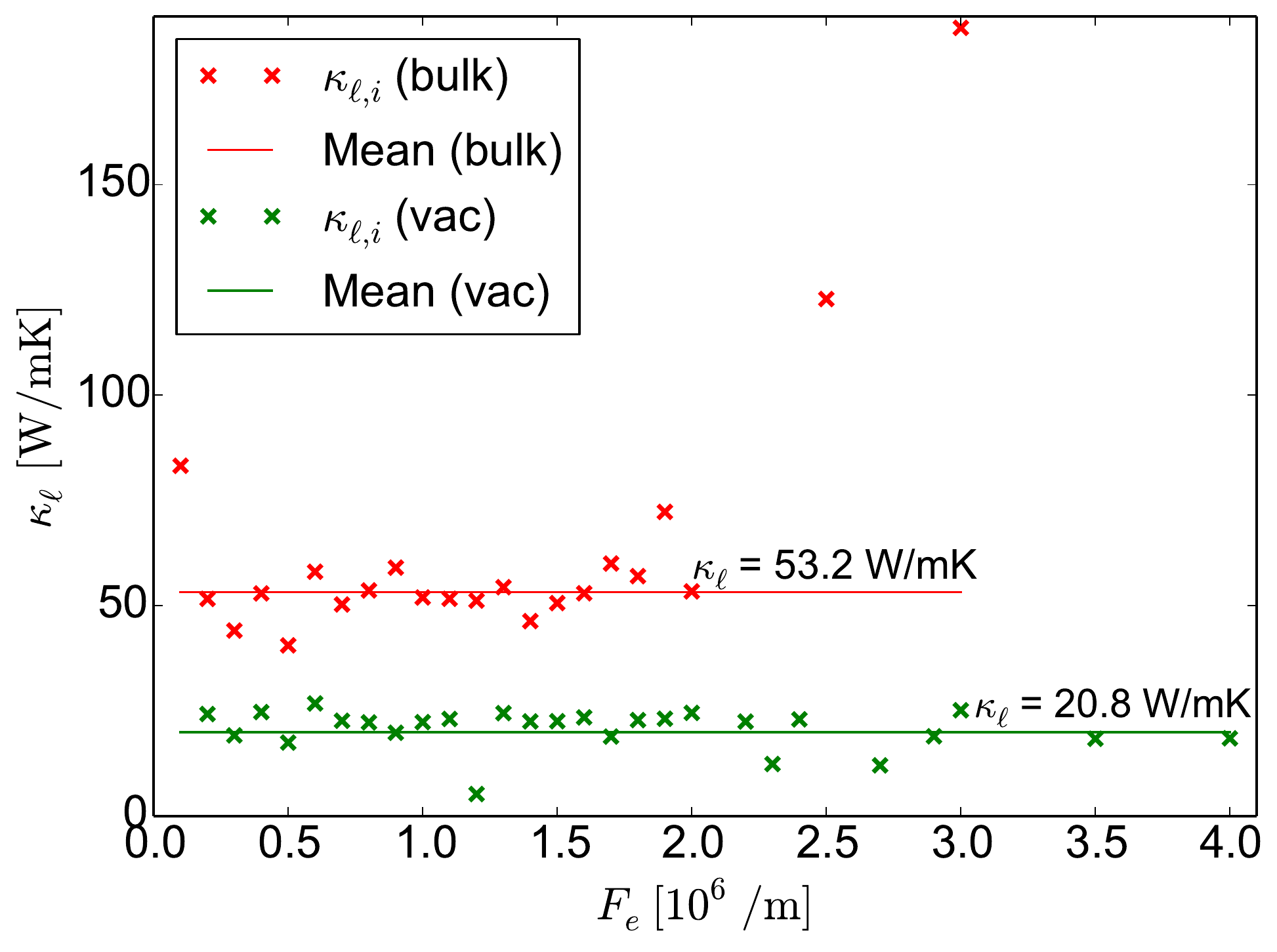}
		\caption{}
		\label{fig:Si_bulk_vac_Fmean}
	\end{subfigure}
\caption{(a) Running average of heat current \textit{\textbf{J}} as a function of $F_{e}$ in bulk (red) and defected Si with five randomly distributed vacancies (green). The thermal conductivity is obtained via the gradient method. (b) $\kappa_\ell$ as a function of $F_{e}$ used to obtain the thermal conductivity via the mean method for both bulk and vacancy Si.}
\label{fig:Si_bulk_vac_kapa_HNEMD}
\end{figure}

From {\bf the HNEMD method} the thermal conductivity was obtained through  Eq.~\eqref{Eqn:J_kFe}. A time-independent external perturbation field $\boldsymbol{F}_e$ was applied in the \textit{x} direction. The calculations were carried out for a total of 22 $F_{e}$ values ranging from 1$\times10^{-5}$ to 3$\times10^{-4}$~(\AA$^{-1}$) at 1000~K for 5$\times10^6$ time steps each of duration of 0.55~fs  and running averages of the heat currents were collected for each simulation. Fig.~\ref{fig:Si_bulk_JvsT} shows the behavior of the running average of the heat current with the external perturbation field, $F_{e}=8 \times 10^{-5}~\mathrm{\AA^{-1}}$. As can be seen in the figure, the initial fluctuations of heat current are very high and they stabilize at $\approx 2000$~ps. Consequently, we choose to cut off sampling $\langle \boldsymbol{J} \rangle$ at $t=2750$~ps. This lead to a bulk thermal conductivity of 53.4~W/(mK) at 1000~K, see Fig.~\ref{fig:Si_bulk_vac_JvsF}. Cutting the sampling at $t=2400$~ps lead to $\kappa_\ell=52.6$~W/(mK), underlining that the results are well converged.

We have considered two methods for calculating $\kappa_\ell$ from $\langle \boldsymbol{J}(t)\rangle$. In the gradient method the slope of a least-squares fit of $\left \langle J_x(t) \right \rangle/VT$ versus $F_{e}$ is identified as the thermal conductivity, Eq.~\eqref{Eqn:J_kFe}. We assume that the intercept is zero.
In the mean method $\kappa_\ell$ is calculated by averaging the $\kappa_{\ell}$'s obtained from several individual runs with varying $F_e$. For bulk-Si, the values of thermal conductivity calculated by the gradient and mean methods are $\kappa_{\ell}$ = 53.4 $\pm$ 1.9 W/(m K) Fig.~\ref{fig:Si_bulk_vac_JvsF} and 53.2  $\pm$ 6.7 W/(m K) Fig.~\ref{fig:Si_bulk_vac_Fmean}, respectively. The values are in good internal agreement the GK calculations as well as the earlier HNEMD calculations \cite{mandadapu2009homogeneous}. 

One observation is that the determination of the range where $\kappa_{\ell}$ depends linearly on $F_{e}$ is important. For the bulk-Si runs, it can be seen in Fig.~\ref{fig:Si_bulk_vac_JvsF} that the values of $\kappa_{\ell}$ deviate strongly from a linear behavior for $F_{e} > 2 \times 10^{-5}$~\AA$^{-1}$. The other possible shortcoming of the HNEMD method is that it is inefficient for very small values of $F_{e}$. At very small values of $F_{e}$ all the three components of $\boldsymbol{J}$ are almost equal suggesting that the system is still in equilibrium.  Due to this, the estimation of the ratio $\langle J(t) \rangle/TF_{e}$ becomes very difficult, as $\left \langle J_x(t) \right \rangle$ approaches zero for these values. Hence, it is crucial to determine a range of $F_{e}$ that is large enough to obtain reasonable values of $\left \langle J_x(t) \right \rangle/TF_{e}$ and small enough for the system to be in the linear non-equilibrium range \cite{mandadapu2009homogeneous}. A fine scan over different values of $F_{e}$ is required to correctly estimate $\kappa_\ell$.

\subsection{Influence of vacancies on $\kappa_\ell$ of Si}
\label{subsec:vacancies}

In order to compare the performance of  GK and HNEMD methods in defected structures, we carried out $\kappa_\ell$ calculations for 1 to 10 (0.06-0.57 atomic \%) randomly distributed vacancies in the $6\times6\times6$ Si supercell (1728 atoms in bulk) at 1000~K. To minimize the interaction of vacancies among themselves, the vacancies were distributed in such a manner that no two vacancies were within the second nearest neighbor distance of one another.

\begin{figure}
\centering
\begin{subfigure}{.5\textwidth}
  \centering
  \includegraphics[width=1.009999\linewidth]{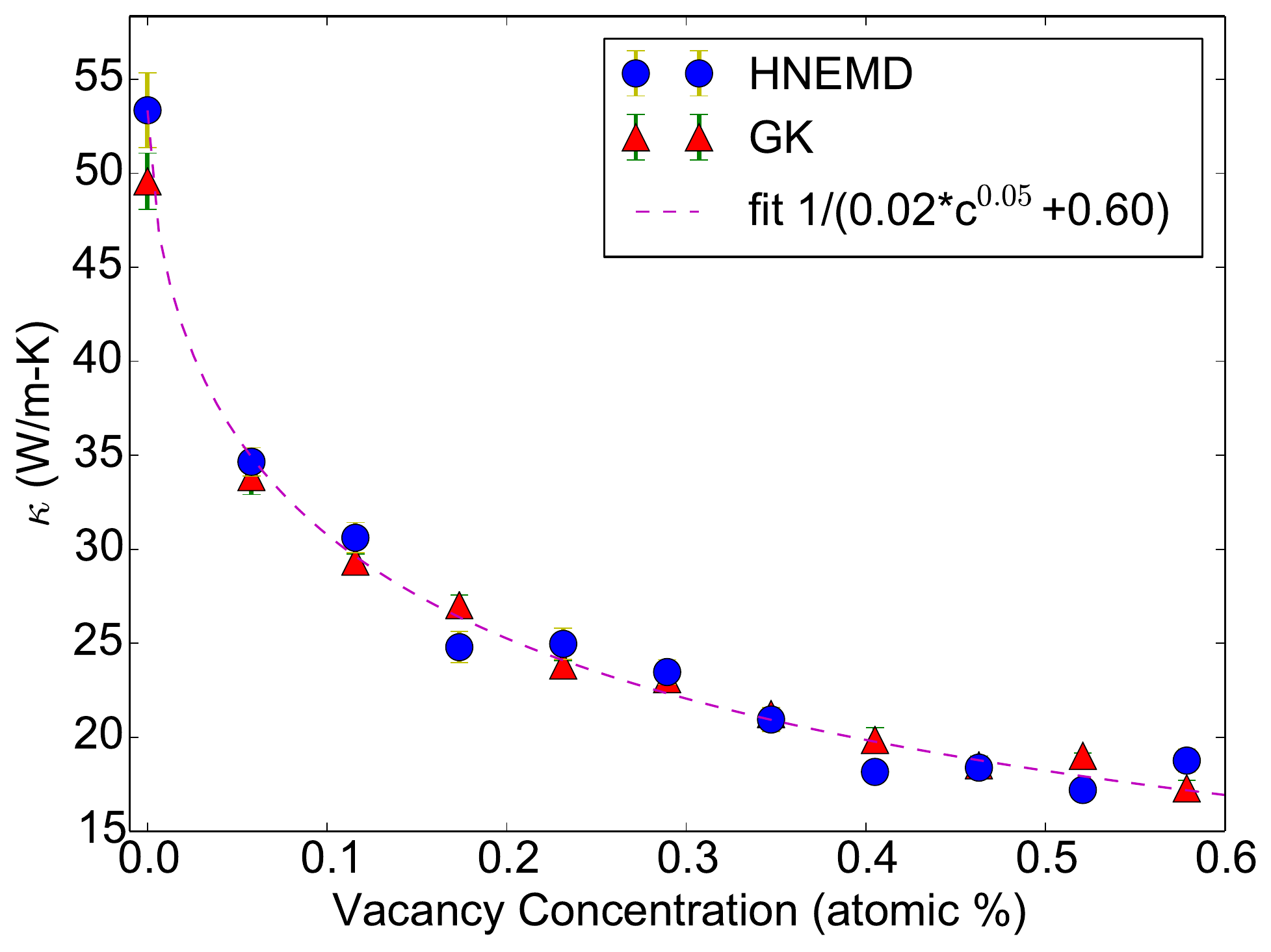}
  \caption{}
  \label{fig:GK_HNEMD_vac}
\end{subfigure}%
\begin{subfigure}{.5\textwidth}
  \centering
  \includegraphics[width=1.\linewidth]{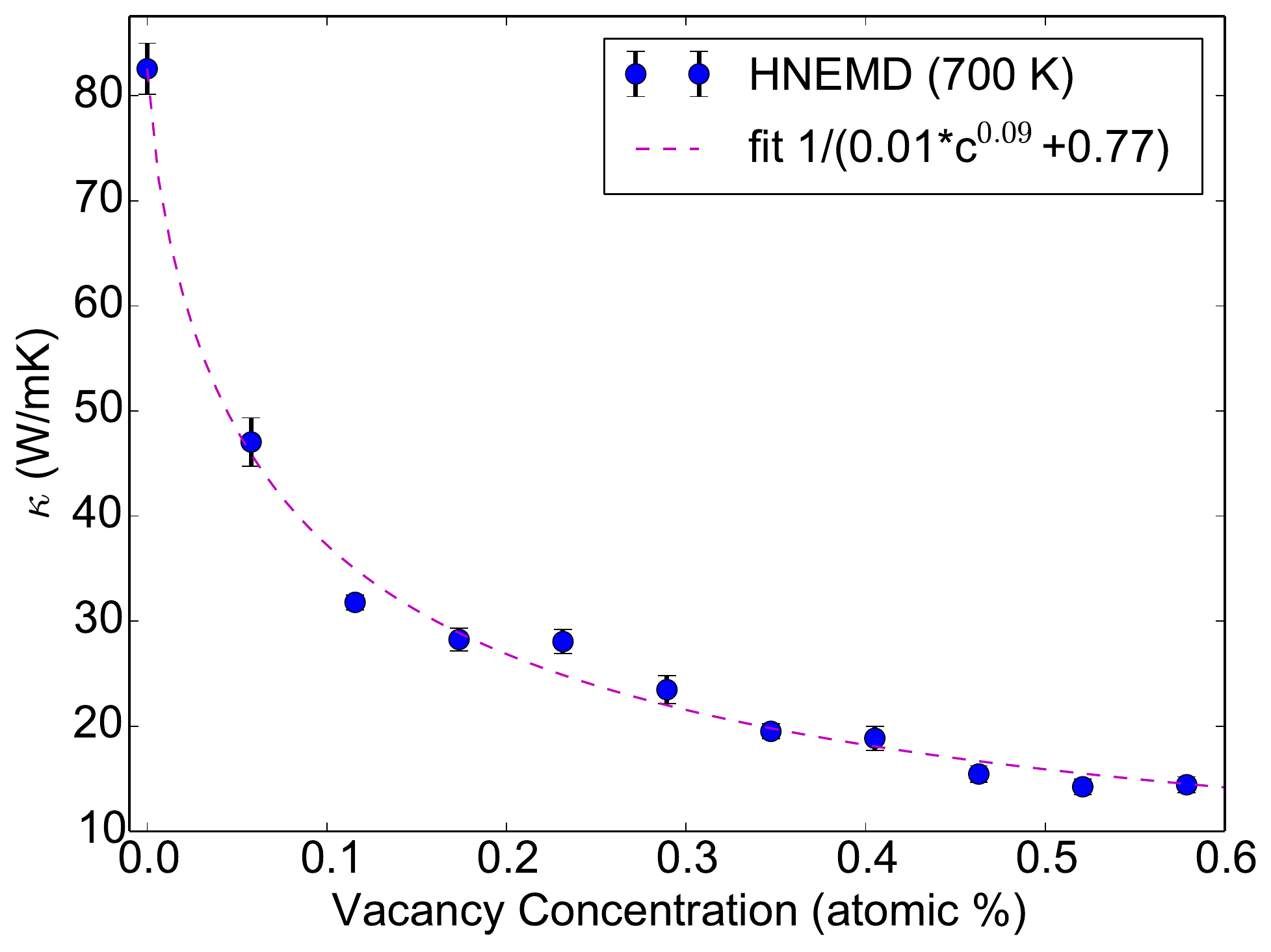}
  \caption{}
  \label{fig:K_vac_700K}
\end{subfigure}
\caption[Kapa vs vacancy count]{(a) Comparison of GK and HNEMD methods with regards to $\kappa_\ell$ vs vacancy concentration for Si at 1000~K. (b) $\kappa_\ell$ vs vacancy concentration for Si at 700~K.}
\label{fig:HNEMD_vac}
\end{figure}

Although, both the methods agree well in terms of thermal conductivity predictions for defect structures, Fig.~\ref{fig:GK_HNEMD_vac}, we would like to point out that for the defected structures the calculation of $\kappa_\ell$ becomes easier with HNEMD method. This is illustrated as follows. In Sec.~\ref{subsec:bulk} we saw that one of the shortcomings of the HNEMD method is the difficulty in determining the linear range of $\left \langle J_x(t) \right \rangle/VT$ vs $F_{e}$. However, in our calculations for defected Si, we observed that as the vacancy concentration increases $\left \langle J_x(t) \right \rangle/VT$ vs $F_{e}$ remains linear for higher and higher values of $F_{e}$. This can be observed in  Fig.~\ref{fig:Si_bulk_vac_JvsF} which shows a comparison of the linear regimes in bulk and defected Si. It can be clearly seen in Fig.~\ref{fig:Si_bulk_vac_JvsF} that one has to do a fine scan over $F_{e}$ values in case of bulk to find out the exact linear range. The relation between $\langle J_x(t) \rangle/VT$ and $F_{e}$ becomes non-linear already around $F_{e} = 2\times10^{-5}$~\AA$^{-1}$, whereas in case of defect structure the plot is still linear for values as high as $40\times10^{-5}  \mathrm{\AA^{-1}}$. This behaviour is also illustrated in Fig.~\ref{fig:Si_bulk_vac_Fmean} where, for bulk the $\kappa_\ell$  values deviate for higher $F_{e}$ whereas for the defected structure they are stable. Hence, in case of defected structures one can obtain the linear range and therefore $\kappa_\ell$ using fewer values of $F_{e}$ thereby reducing the computational cost. 

We found an inverse power-law like decay of $\kappa_\ell$ against the vacancy concentration, $c$, for both GK and HNEMD methods confirming a good agreement between the two methods, Fig.~\ref{fig:GK_HNEMD_vac}. Following the discussion above, we performed $\kappa_\ell$ calculations at 700~K using only HNEMD method. 
An inverse power-law decay for $\kappa_\ell$ vs $c$ was also observed at 700~K, as shown in Fig.~\ref{fig:K_vac_700K}. We also tried to fit  exponential functions to $\kappa_\ell$'s but the standard deviation for the fit values was orders of magnitude higher than the power-law fit. Hence we can establish that the thermal conductivity falls according to inverse power-law with increasing vacancy concentration in crystalline Si. Our results agree well the literature \cite{lee2011effects, wang2014atomistic, shahraki2015effects}.

 \section{Conclusions}
 \label{sec:conclusions}
 We have done a detailed study of the GK and HNEMD methods for calculating $\kappa_\ell$ in bulk and defected Si and highlighted the advantages and disadvantages of each method. We have discussed the underlying parameters of the two methods and shown that by a judicious choice both the methods performed equally well for $\kappa_\ell$ calculations in bulk Si. The GK method in the above fashion produces very accurate values of thermal conductivity within statistical uncertainties. With the GK method the entire lattice thermal conductivity tensor can be calculated from one simulation, unlike the HNEMD method which necessitates several simulations in each direction to achieve the same. At the same time the GK method can suffer from averaging and truncation errors which makes it more cumbersome than the HNEMD method. Both the GK and the HNEMD method require several independent simulations to get a $\kappa_\ell$ with a low uncertainty. These simulations in the respective methods are not interdependent and can be run in parallel. The HNEMD method has an advantage over the GK method as it generates lesser statistical errors and reduces the necessary computation time by combining the elements of both equilibrium and non-equilibrium MD simulations. Specially in case of defect structures HNEMD can be advantageous.

\section{Acknowledgements}
\label{sec:acknow}
We acknowledge Dr.~Mandadapu and Prof.\ Dr.\ Papadopoulos for the helpful discussions and the HNEMD code, and the European Union’s Horizon 2020 Research and Innovation Programme, grant number 645776 (ALMA).

\section*{References}
\bibliographystyle{iopart-num} %unsrtnatMSMSE
\bibliography{msmsebib}

\end{document}